\documentclass[aps,pra,twocolumn,amsmath,amssymb,groupedaddress]{revtex4-1}

\usepackage{graphicx}
\usepackage{subfigure}
\usepackage{graphics}

\begin{document}

%\preprint{}

\title{Rogue solitons in Heisenberg spin chain}


\author{Aritra K. Mukhopadhyay}
\email{aritra1910@gmail.com}
%\thanks{}
\affiliation{Indian Institute of Science Education and Research (IISER) Kolkata, Mohanpur - 741 246, India.}

\author{Vivek M. Vyas}
\email{physics.vivek@gmail.com}
%\thanks{}
\affiliation{Institute of Mathematical Sciences, IV Cross Road, CIT Campus, Taramani, Chennai 600 113, India.}

\author{Prasanta K. Panigrahi}
\email{pprasanta@iiserkol.ac.in}
%\thanks{}
\affiliation{Indian Institute of Science Education and Research (IISER) Kolkata, Mohanpur - 741 246, India.}


\date{\today}

\begin{abstract}
Following the connection of the non-linear Schr\"{o}dinger equation with the continuum Heisenberg spin chain, we find the rogue soliton equivalent in the spin system. The breathers are also mapped to the corresponding space or time localized oscillatory modes, through the moving curve analogy. The spatio-temporal evolution of the curvature and torsion of the curve, underlying these dynamical systems, are explicated to illustrate the localization property of the rogue waves. 
\end{abstract}

%\keywords{}

\maketitle

%\section{}
%\subsection{}
%\subsubsection{}

\section{Introduction}
In a remarkable paper \cite{lakh3}, Lakshmanan established the exact connection between the continuum Heisenberg spin chain in one dimension and non-linear Schr\"{o}dinger equation (NLSE). This mapping made use of moving curve dynamics, which has been earlier used to find solitons on a vortex filament \cite{hasi}. The same connection also extends to many other integrable systems admitting soliton solution. The Manakov system, described by two moving interacting curves \cite{kost}, modified Korteweg-de Vries \cite{nak}, sine-Gordon \cite{lamb}, Ishimori and Myrzakulov system \cite{lakh4} are among many other models connected to moving curve dynamics. The subject of moving curves and surfaces has received significant attention, since the underlying geometry can provide deep insight into the dynamics of these diverse physical systems \cite{lakh1}. Other examples of physical interest include dynamics of filament vortices in ordinary and super fluids \cite{hasi}, spin systems \cite{lakh3}, phases in classical optics \cite{kug} and many systems encountered in physics of soft matter \cite{bird}. The evolution of these curves gets connected to the non-linear evolution equations through the geometrical properties of the curves, such as curvature and torsion. The exact solvability of the equations describing moving curves enables one to study these physical systems, often intractable otherwise.

\paragraph*{}As is well known, the integrable NLSE has a host of solutions, which finds applications in various areas. It is an integrable system, possessing solitary wave solutions, that arise due to the interplay of the dispersive and non-linear effects in a medium. The observed dark, bright and grey solitons are different limiting cases of the non-linear cnoidal waves \cite{sulem}. The well-known envelope solitons of the NLSE have been observed in many different systems including plasma, optical fiber and cold atoms \cite{env1,env2,env3}. NLSE also allows for ``breather" solutions, Akhmediev \cite{akh1,akh2} and the Kuznetsov-Ma breathers \cite{kuz}, in which energy concentration is localized and oscillatory. Interestingly, there exists another class of highly localized solution, the Peregrine breather, which is obtained as the limiting case of the space-periodic Akhmediev breather and the time-periodic Kuznetsov-Ma breather, when the period tends to infinity. It accurately models `rogue waves' in oceans, which are highly localized in the spatio-temporal domain, appearing out of nowhere in the open seas, attaining great amplitudes and then promptly disappearing, causing great devastation. Initially discovered in 1983 by Peregrine \cite{per}, the Peregrine breather was the first clear and mathematically precise model of this rogue wave phenomenon. Being a rare event, the experimental observation of this type of solitons remained a difficult task. It was only in 2010 that the `rouge wave' was finally detected in optical fibers \cite{po}. Later, it was also observed in the waves generated in the multi-component plasma \cite{pp} and in an experimental water tank \cite{pw}. Recent works have also shown the existence of controlled giant rogue waves in non-linear fiber optics \cite{pkp1} and dissipative optical rogue waves in mode-locked fiber lasers \cite{dis}.

\paragraph*{} Here, we make use of the established connection between one dimensional continuum Heisenberg spin chain and the NLSE to find the rogue wave equivalent in this 1D magnetic system. The moving curve analogy reveals the localized spatio-temporal evolution of the curvature and torsion, connecting the rogue wave solution to its spin chain excitation. The breathers are also mapped to the corresponding space or time localized oscillatory modes in the spin system, through this correspondence. We conclude by highlighting some key differences between the rogue wave spin excitations and the previously found envelope soliton modes in the spin system. Directions of future study are also indicated.

\section{Spin rogue waves in 1D continuum spin chain}
The Frenet-Serret equations of space curves establish the link between the dynamics of non-linear systems mentioned earlier and geometry. Lakshmanan established an exact map of the continuum limit of the one dimensional classical spins with nearest neighbour Heisenberg interactions, governed by the equation of motion $\frac{\partial S(x,t)}{\partial t}= S(x,t)\frac{\partial^2 S(x,t)}{\partial x^2}$,  to the non-linear Schr\"{o}dinger equation \cite{lakh2}. Identifying the unit spin vector with the tangent to a helical curve having a curvature $\kappa(x,t)$ and torsion $\tau (x,t)$, the energy and momentum density can be expressed in terms of the torsion and curvature \cite{lakh3}: $\mathcal{E}_{spin} (x,t)= \frac{1}{2}|\frac{\partial S(x,t)}{\partial x}|^2=\frac{1}{2}\kappa^2(x,t)$, $\mathcal{P}_{spin} (x,t)= S\frac{\partial S(x,t)}{\partial x} \frac{\partial^2 S(x,t)}{\partial x^2}=\kappa^2(x,t)\tau (x,t)$.

\paragraph*{}
For certain choice of dynamics of the curve, the evolution of the torsion and curvature parameters can be obtained as the solution of NLSE : $i\frac{\partial \psi}{\partial t} + \frac{1}{2}\frac{\partial^2 \psi}{\partial x^2} + |\psi|^2 \psi=0$. $\psi$ is related to the torsion and curvature by, $\psi(x,t)=\kappa(x,t)\exp\left\lbrace i\int^x\tau(x,t)dx \right\rbrace$. As the NLSE is exactly solvable, the continuum spin system is also an exactly solvable system. For the Peregrine breather,
\begin{equation}
\psi(x,t)=\left[ 1-\frac{4(1+2it)}{1+4x^2 +4t^2}\right] e^{it}
\end{equation}
the curvature is obtained as $ \kappa(x,t)=\sqrt{\left(1 - \frac{4}{1 + 4 t^2 + 4 x^2}\right)^2 + \left( \frac{8 t}{1 + 4 t^2 + 4 x^2}\right)^2}$. The torsion can also be calculated: $\tau(x,t)$ = $\frac{-64tx}{16t^4 + (3-4x^2)^2 +8t^2 (5+4x^2)}$. As is evident, both these quantities have a localized Lorentzian character. We also note that that the curvature and energy is invariant under parity and time-reversal transformations, whereas torsion and momentum is invariant only under the joint transformation. This is to be expected as the torsion depends on the left or right handed coordinate system.

\begin{figure*}
  \subfigure[\small \sl Curvature]{\includegraphics[width=.4\linewidth]{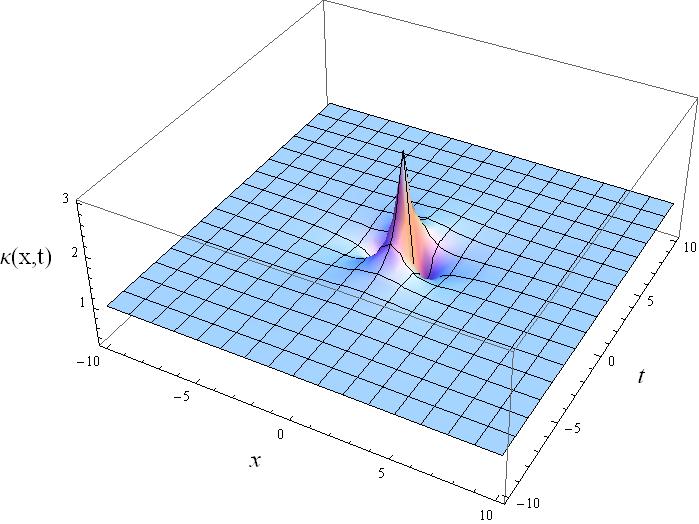} }\hfill
  \subfigure[\small \sl Torsion]{\includegraphics[width=.4\linewidth]{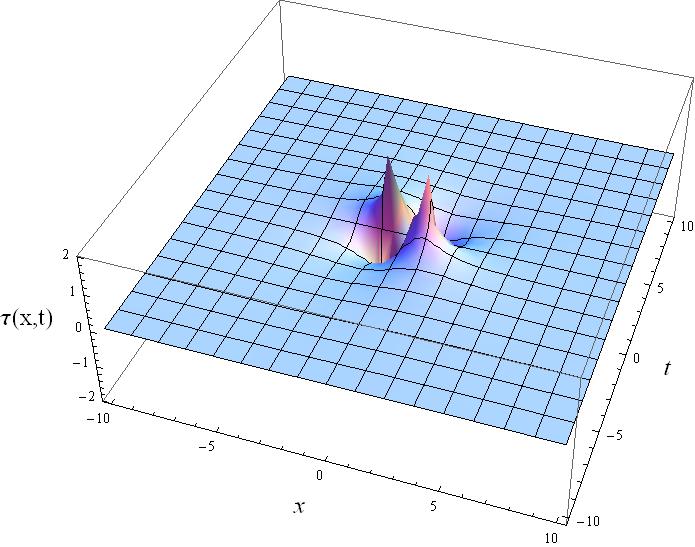}}\hfill
  \caption{\small \sl Curvature and torsion of the helical curve (as a function of space and time) which has the spin vector S(x,t) as its tangent. }
\end{figure*}

\begin{figure}
  \includegraphics[scale=0.3]{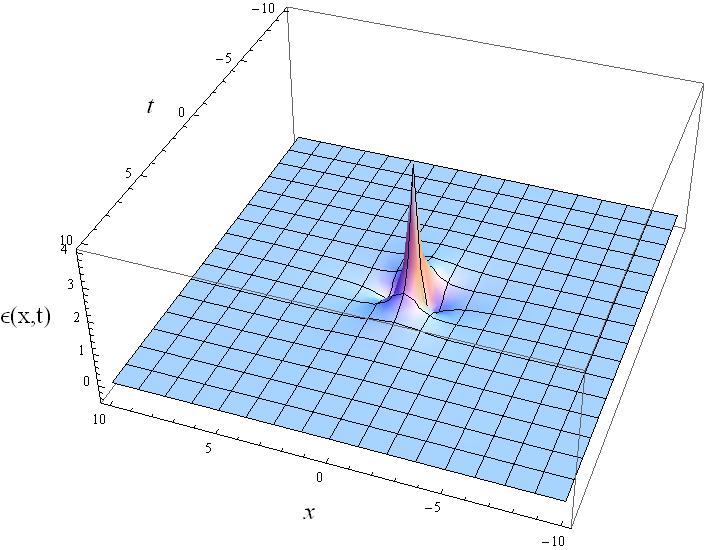}\hfill
  \caption{\small \sl Spatio-temporally localized energy density in spin chain.}
\end{figure}

\begin{figure*}
  \subfigure{\small \sl \includegraphics[width=.4\linewidth]{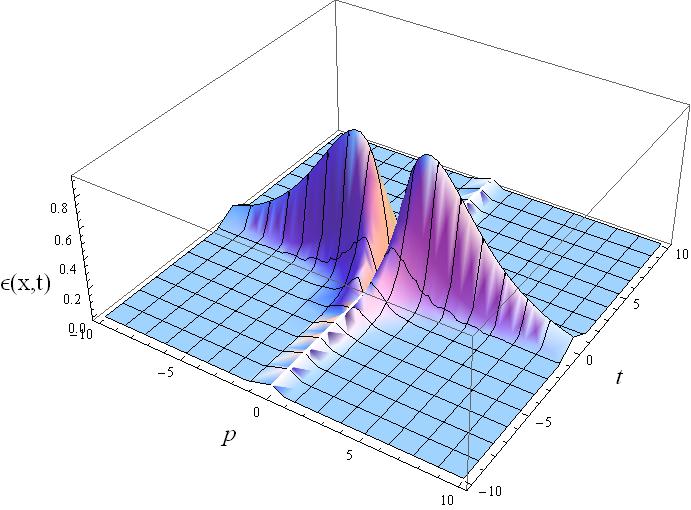}}\hfill
  \subfigure{\small \sl \includegraphics[width=.3\linewidth]{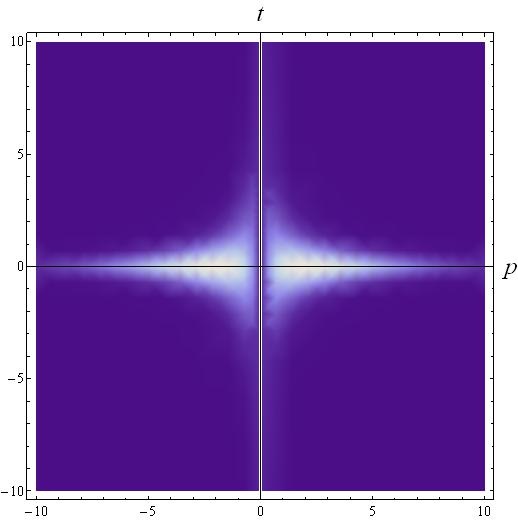}}\hfill
    
  \caption{\small \sl [a]3D plot and [b]density plot showing the energy density of the spin chain in the fourier space}
\end{figure*}

\paragraph*{}
The momentum density for the spin system is given by
\begin{equation}
\mathcal{P}_{spin} (x,t)=\kappa^2(x,t)\tau (x,t)=\frac{-64tx}{(1+4t^2 +4x^2)^2}
\end{equation}
and the energy density,
\begin{widetext}
\begin{equation}
\mathcal{E}_{spin} (x,t)=\frac{1}{2}\left[\left(1 - \frac{4}{1 + 4 t^2 + 4 x^2}\right)^2 
+ \left( \frac{8 t}{1 + 4 t^2 + 4 x^2}\right)^2\right]
\end{equation}
\end{widetext}
After subtracting the background contribution from the energy density, one observes that the spin system exhibits a spatio-temporally localized absorption mode. Due to the Lorentzian nature, the energy is largely concentrated at the center of the chain within a width of $\sqrt{1+4t^2}$. As the total energy is zero, which can be explicitly seen by integrating the energy density over space, the spin mode exchanges energy with the background. The energy density peak flattens over time and after a prolonged period, there remains only a constant background. In the fourier domain, the energy density takes the form
\begin{equation}
\mathcal{E}_{spin} (p,t)=\frac{1}{2}(e^{-\frac{|p|}{2 \sqrt{\frac{1}{1+4 t^2}}}} \sqrt{2 \pi } |p|+\sqrt{2 \pi } \delta(p))
\end{equation}
thus showing a condensation and an exponentially decaying mode. The constant background manifests as condensation and the Lorentzian part leads to the exponential decay. To find the time and length scale of the system, we restore the dimensionful parameters
\begin{widetext}
\begin{equation}
\mathcal{E}_{spin} (x,t)=\frac{1}{2}\left[\left(1 - \frac{4}{1 + 4 (\frac{t}{T})^2 + 4 (\frac{x}{X})^2}\right)^2 + \left( \frac{8 (\frac{t}{T})}{1 + 4 (\frac{t}{T})^2 + 4 (\frac{x}{X})^2}\right)^2\right]
\end{equation}
\end{widetext}
Here $X=\frac{\hbar}{\sqrt{mg}}$ and $T=\frac{2\hbar}{g a^2 J}$, \emph{g} is the coupling strength appearing in the NLSE, \emph{m} is the mass, \emph{a} is the lattice constant between neighboring sites and \emph{J} is the spin coupling strength.

\section{Spin breathers in 1D continuum spin chain}
To understand the rogue wave modes of the spin chain from very general perspective, we study the spin equivalent of the Akhmediev breather,
\begin{equation}\label{breath}
\psi=\frac{(1 - 4 a) \cosh[b t] + \sqrt{2 a} \cos[\Omega x] + i b \sinh[b t]} {\sqrt{2 a} \cos[\Omega x] - \cosh[b t]} e^{it}
\end{equation}

where $\Omega$ is the dimensionless modulation frequency, $a=\frac{1}{2}(1-\frac{\Omega^2}{4})$ and $b=\sqrt{8a(1-2a)}$. In two different limits of the breather solution one can obtain the continuous wave ($a\longrightarrow 0$) and the Peregrine solution ($a\longrightarrow \frac{1}{2}$). In the polar form, $\psi=r(x,t)e^{i\alpha (x,t)}$. One finds, $r^2(x,t) = A^2(x,t) +B^2(x,t) \label{akh1}$, $\alpha (x,t)= \frac{1}{2} \tan^{-1} \left(\frac{-2A(x,t)B(x,t)}{A(x,t)^2 - B(x,t)^2}\right)-t$ \label{2} , $A(x,t)= \frac{(1 - 4 a) \cosh[b t] + \sqrt{2 a} \cos[\Omega x] }{\sqrt{2 a} \cos[\Omega x] - \cosh[b t]}$ and $B(x,t)= \frac{b \sinh[b t]}{\sqrt{2 a} \cos[\Omega x] - \cosh[b t]}$.

\paragraph*{} Proceeding as before one finds $\kappa(x,t)=r(x,t)$ and the torsion, $\tau(x,t)= \frac{\partial \alpha}{\partial x}$. The energy density of the spin system is then obtained as
$\tilde{\mathcal{E}}_{spin} (x,t)=\frac{1}{2}r^2(x,t)=\frac{1}{2} \frac{[(1 - 4 a) \cosh[b t] + \sqrt{2 a} \cos[\Omega x]]^2 + [b \sinh[b t]]^2}{(\sqrt{2 a} \cos[\Omega x] - \cosh[b t])^2}
$.

Evidently one observes a periodic behavior in the energy density profile. The limit of $a\longrightarrow 0$ yields the continuous background. A slight variation in the parameter \emph{a} triggers oscillatory excitation modes in the spin chain. With the passage of time, the widths of these localized peaks increase and the peak heights decrease. The maximum localization is obtained at t=0. For the sake of clarity, we study this excitation mode at t=0 for various parameter values $a$ after subtracting the constant background. We find that at t=0,
\begin{widetext}
\begin{equation}
\mathcal{E}_{spin} (x,0)=\frac{1}{2}(r^2(x,0)-1)=\frac{1}{2}\left[ \left(\frac{(1 - 4 a) + \sqrt{2 a} \cos[\Omega x] }{\sqrt{2 a} \cos[\Omega x] - 1 }\right)^2-1\right]
\end{equation}
\end{widetext}
The width of the peaks, measured by the distance between the two adjacent zeros, is found as $w=\frac{2}{\Omega} cos^{-1} \sqrt{2a}=\frac{1}{\sqrt{1-2a}} cos^{-1} \sqrt{2a}$, as the two consecutive zeros occur at $x=\pm \frac{1}{2\sqrt{1-2a}} cos^{-1} \sqrt{2a},\ \pm \frac{\pi}{\sqrt{1-2a}}\pm \frac{1}{2\sqrt{1-2a}} cos^{-1} \sqrt{2a}\ $ etc.

\begin{figure*}
  \subfigure[\small for a=0.4]{\includegraphics[width=.4\linewidth]{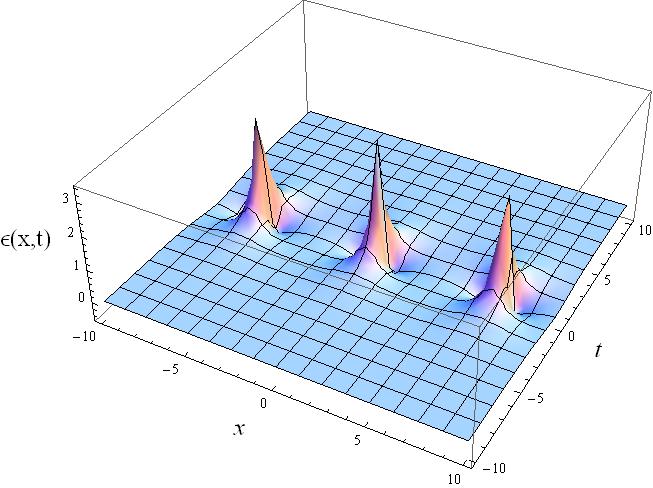} }\hfill
  \subfigure[\small for a=0.4999]{\includegraphics[width=.4\linewidth]{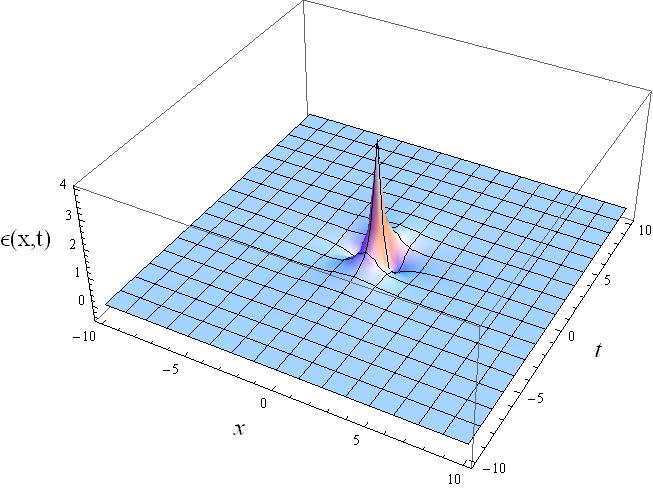}}\hfill
  \caption{\small \sl Periodic and oscillatory energy density of the continuum spin chain calculated using the map between Akhmediev breather solution and the spin chain \label{spinakh2}}
\end{figure*}

\begin{figure}
\includegraphics[scale=0.38]{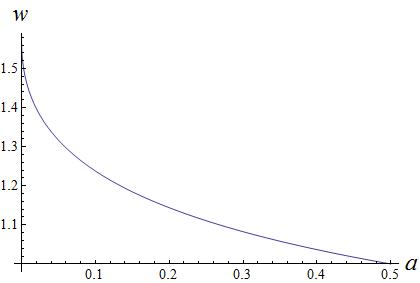} 
\caption{\small \sl Variation of energy density peak width 'w' with the parameter 'a'\label{wvsa}}  
\end{figure}

It is noted that the peak width decreases as the parameter \emph{a} increases, as is clearly seen in Figure \ref{wvsa}. It is also found that for $a\longrightarrow 0.5$, the width approaches unity, which corresponds to excitation mode of the spin chain obtained for the Peregrine breather solution in the previous section. It is evident that the continuum spin chain exhibits periodic energy absorption modes and exchanges energy with the background. The integral of the energy density over a whole period is zero. Considering the peak at $x=0$, we find that the energy absorbed between the region $\left[-\frac{1}{\Omega} cos^{-1} \sqrt{2a}, \frac{1}{\Omega} cos^{-1} \sqrt{2a}\right]$ is exactly equal to the energy exchanged with the background in the intervals $\left[ -\frac{\pi}{\Omega},-\frac{1}{\Omega} cos^{-1} \sqrt{2a} \right]$ and  $\left[ \frac{1}{\Omega} cos^{-1} \sqrt{2a}, \frac{\pi}{\Omega} \right]$, since $|E_1| + |E_2| =E_3$ and  $E_1+E_2+E_3=0$. Here
$E_1= \int_{-\frac{\pi}{\Omega}}^{-\frac{1}{\Omega} cos^{-1} \sqrt{2a}} \mathcal{E}_{spin} (x,t)dx$, $E_2= \int_{\frac{1}{\Omega} cos^{-1} \sqrt{2a}}^{\frac{\pi}{\Omega}} \mathcal{E}_{spin} (x,t)dx$ and $E_3= \int_{-\frac{1}{\Omega} cos^{-1} \sqrt{2a} }^{\frac{1}{\Omega} cos^{-1} \sqrt{2a}} \mathcal{E}_{spin} (x,t)dx$. The same analysis can be repeated for the time periodic Kuznetsov-Ma breather which would yield spin waves localized in space and periodic in time.

\section{Conclusion}
In conclusion, the continuum limit of the one-dimensional Heisenberg spin system admits spin wave equivalent of the Akhmediev and Ma breathers. It further admits highly localized rogue wave equivalent of spin excitation which asymptotically shows constant magnetization. It is found that the spatial and temporal widths of the spin waves can be controlled by changing the coupling parameters. The space curve route used here to establish the connection between NLSE and spin system illustrates the curvature and torsion configurations behind these excitation. The fact that skyrmions have been experimentally realized in spin systems \cite{sky} gives us hope that the present excitation may find experimental verification. In future, we would like to investigate such systems for variable coefficients \cite{dai} which may enable its coherent control and amplification.

\bibliography{final_spin}

\end{document}